\documentclass[superscriptaddress,letterpaper, aps, amssymb, preprint, amsmath, 12pt]{revtex4-1}

\usepackage[justification=raggedright,singlelinecheck=false,labelfont=bf,labelsep=period]{caption,subcaption}

\usepackage{braket}
\usepackage{siunitx}
\usepackage{comment}
\usepackage{nameref}

\newcommand{\velocity}{m s$^{-1}$}

\newcommand{\wmk}{Wm$^{-1}$K$^{-1}$}

\newcommand{\SM}{Supporting Information}
\usepackage{scalerel}
\usepackage{tikz}
\usetikzlibrary{svg.path}
\definecolor{orcidlogocol}{HTML}{A6CE39}
\tikzset{
  orcidlogo/.pic={
    \fill[orcidlogocol] svg{M256,128c0,70.7-57.3,128-128,128C57.3,256,0,198.7,0,128C0,57.3,57.3,0,128,0C198.7,0,256,57.3,256,128z};
    \fill[white] svg{M86.3,186.2H70.9V79.1h15.4v48.4V186.2z}
                 svg{M108.9,79.1h41.6c39.6,0,57,28.3,57,53.6c0,27.5-21.5,53.6-56.8,53.6h-41.8V79.1z M124.3,172.4h24.5c34.9,0,42.9-26.5,42.9-39.7c0-21.5-13.7-39.7-43.7-39.7h-23.7V172.4z}
                 svg{M88.7,56.8c0,5.5-4.5,10.1-10.1,10.1c-5.6,0-10.1-4.6-10.1-10.1c0-5.6,4.5-10.1,10.1-10.1C84.2,46.7,88.7,51.3,88.7,56.8z};}}
\newcommand\orcidicon[1]{\href{https://orcid.org/#1}{\mbox{\scalerel*{
\begin{tikzpicture}[yscale=-1,transform shape]
\pic{orcidlogo};
\end{tikzpicture}
}{|}}}}

\usepackage{etoolbox}
\makeatletter
\pretocmd\frontmatter@thefootnote{\color{blue}}{}{}
\usepackage{dcolumn}
\usepackage{titlesec}
\titlespacing*{\section}
{0pt}{4ex}{1.3ex}
\titlespacing*{\subsection}
{0pt}{4ex}{0ex}
\titleformat{\section}
  {\bfseries\MakeUppercase}{\thesection\raggedright}{1em}{}
  \titleformat{\subsection}
 {\bfseries}{\thesubsection\raggedright}{1em}{}
\usepackage{caption}
\usepackage{xstring}
\usepackage{xspace}
\usepackage[utf8]{inputenc}
\usepackage{natbib}
\usepackage{gensymb}
\usepackage{mathtools}

\usepackage{float}
\usepackage{graphicx}

\usepackage{url}
\usepackage[utf8]{inputenc}
\usepackage[T1]{fontenc}
\usepackage{textcomp}
\usepackage{gensymb}
\usepackage[colorlinks,citecolor=red,urlcolor=blue,bookmarks=false,hypertexnames=true]{hyperref}
\usepackage{booktabs}
\usepackage{multirow}
\usepackage{textcomp}
\usepackage{graphicx}

\usepackage{cleveref}
\renewcommand{\eqref}[1]{\ref{#1}}
\crefname{equation}{}{}
\makeatletter
\crefformat{equation}{#2#1#3}
\crefmultiformat{equation}{#2#1#3}{,~#2#1#3}{,~#2#1#3}
\makeatother

\begin{document}
\title{Thermal transport and the impact of hydrogen adsorption in Linde Type A zeolitic imidazolate frameworks}
\author{Hyunseok Oh~\orcidicon{0009-0001-2571-8209}}
\affiliation{%
Department of Mechanical Engineering, Seoul National University, Seoul 08826, Republic of Korea}%
\author{Taeyong Kim~\orcidicon{0000-0003-2452-1065} }
 \email{tkkim@snu.ac.kr}
\affiliation{%
Department of Mechanical Engineering, Seoul National University, Seoul 08826, Republic of Korea}%
\affiliation{
Institute of Advanced Machines and Design, Seoul National University, Seoul 08826, Republic of Korea
}
\affiliation{%
Inter-University Semiconductor Research Center, Seoul National University, Seoul 08826, Republic of Korea}%
\date{\today}

\begin{abstract}
Thermal transport in metal-organic frameworks (MOFs) is of practical interest in diverse applications such as gas storage and separations, since insufficient heat dissipation can lead to detrimental effects. Despite investigations, influence of molecular infiltration on the heat transport remains unclear in many of MOFs due to poor understanding of mechanisms governing heat conductions. Here, we report molecular dynamics investigations of thermal transport properties in zeolitic imidazolate frameworks (ZIFs). We investigated Linde Type A topological ZIFs (ZIF-lta) exhibiting exceptionally low thermal conductivity with unusual trend of temperature dependence deviating from many crystalline materials, despite long-range crystalline order in them. We demonstrate that heat is predominantly carried by phonons with mean free paths comparable to their wavelengths, analogous to diffusons in amorphous solids owing to strong anharmonicity caused by complexity of unit cell consisting of a large number of metal centers. We further show that adsorbed hydrogen molecules increase thermal conductivity of ZIFs, mainly contributed by additional vibrational modes, as a result of gas-gas or gas-framework interactions. Our work advances fundamental understanding into the thermal transport in MOFs and suggests a means to engineer heat conduction via gas infiltrations.
\end{abstract}
\maketitle

\section{Introduction}

Metal-organic frameworks (MOFs), inorganic-organic hybrid polymers, are of practical importance in diverse applications including gas adsorption/separation~\cite{rosi2003hydrogen, yu2017co2, mason2014evaluating}, catalysis~\cite{wang2019state}, water harvesting~\cite{kim2017water}, and energy storage~\cite{raya2020pump} due to controllable porosity and high thermal stability. Largely through synthetic flexibility owing to multiple possible combinations of metal nodes and organic linkers~\cite{li2022tuning, deng2012large, burigana2025linker}, their properties can be tailored to specific target applications. For instance, their gas uptake capacities can be optimized by rational design of organic linkers, as demonstrated by MOF-210, an isoreticular expansion of MOF-5, which has a high Brunauer-Emmett-Teller surface area exceeding 6200 m$^2$ g$^{-1}$ along with a pore volume of $\sim3.6$ cm$^3$ g$^{-1}$~\cite{furukawa2010uptake}. Conjugated MOFs exhibit intriguing properties such as tunable band structure~\cite{chen2015bandgap}, and broadband optical response~\cite{arora2020broadband}, advantageous for optoelectronic applications. Additionally, their unusual thermo-electric and photo-induced ionic transport offer opportunities in energy conversion applications~\cite{sun2017thermoelectric,wang2024osmotic}.

Prior works have established that MOFs generally exhibit exceptional heat conduction properties deviating from many crystalline solids, even though they possess long-range crystalline order~\cite{zhang2021DFT,huang2007MD, zhou2022origin}. These works have reported that inherent mass difference between metal nodes and organic linkers impedes the heat transport due to strong phonon scattering distinctly occurring at the nodes. Meanwhile, normal mode analysis based on Allen and Feldman theory~\cite{allen1999diffusons} has been adopted to elucidate their unusual temperature dependence of the thermal conductivity, as MOFs are complex crystals with a high degree of atomic disorder. It has been suggested that the vibrational modes with exceedingly short mean free paths (MFPs) comparable to wavelengths, analogous to diffusons in amorphous solids, predominantly carry the heat, leading to unusual temperature dependencies~\cite{huang2007MD}. Recent computations~\cite{zhou2022origin,zhou2021vibrational} and experiments~\cite{sorensen2020glass} on zeolitic imidazolate frameworks (ZIFs) such as ZIF-4 and ZIF-62 have confirmed these observations, concerning the dominant contribution of these modes to the heat conductions.

Impact of molecular adsorption on the thermal transport in MOFs is of another interest, since gas adsorption is exothermic, which requires heat dissipation to maintain adsorption capacity~\cite{wieme2019uptake}. In some cases, interactions between frameworks and adsorbates were observed to enhance anharmonic scattering, thereby reducing the overall thermal conductivity~\cite{babaei2020observation,babaei2016ideal}. For instance, prior studies have reported that thermal conductivity in HKUST-1 can be diminished up to around 50\% at H$_2$O density of $\sim0.02$ g cm$^{-3}$ due to excessive phonon scattering occurring at low frequencies~\cite{babaei2020observation}. At the same time, it has been suggested that the adsorbates exert contrasting effect on the thermal conductivity facilitated by enhanced heat transfer through the adsorbate phase itself or across the adsorbate–framework interface~\cite{yamaguchi2024aniso,cheng2023lta,han2014relation}. For instance, it has been demonstrated that increased thermal conductivity for copper tetrakis(4-carboxyphenyl)porphyrin MOF with the addition of H$_2$O primarily originates from additional heat propagation through the network formed by hydrogen-bonding between H$_2$O molecules, rather than through gas-framework interactions~\cite{yamaguchi2024aniso}. Concerning the impact of gas-framework interactions, in particular, prior studies have suggested that the thermal conductance which impacts overall effective heat flux flowing across the interface between lattice and guest molecules, are affected by multiple factors such as pore size and geometry~\cite{yuan2024glass, cheng2023lta}, inter-atomic potential strength~\cite{yuan2024glass}, and gas diffusion and distributions~\cite{cheng2023lta}. For example, the increase in thermal conductivity upon molecular adsorption in ZIFs was attributed to lattice-gas interactions, which are more pronounced for H$_2$ than C$_2$H$_5$OH due to its higher mass diffusivity~\cite{cheng2023lta}. Overall, these findings highlight the highly system-dependent yet complicated nature of MOF-gas interactions in governing heat transport in these substances.

Here, we report thermal transport properties and impact of hydrogen molecules adsorption using molecular dynamics (MD) simulations in ZIF-lta. We observe that the ZIF-lta exhibits remarkably low thermal conductivity that lacks temperature dependence, despite long-range crystalline order in them. We attribute our observations to the dominance of thermal phonons with mean free paths comparable to their wavelengths, analogous to diffusons in amorphous materials, owing to strong anharmonicity caused by complex unit cell with a large number of metal centers. We also show that hydrogen gas adsorption enhances thermal conductivity, primarily due to extra vibrational modes arising from the gas-gas and gas-framework interactions. Our work provides fundamental insights into thermal transport in MOFs and suggests a means to manipulate heat conduction through gas infiltrations.

\section{Methods}\label{sec:method}
We studied thermal transport properties of ZIF-lta using equilibrium MD (EMD) simulations implemented in Large-scale Atomic/Molecular Massively Parallel Simulator (LAMMPS). In all simulations, the ZIF-FF~\cite{weng2019zifff} was employed to describe inter-atomic interactions within the frameworks, while the Buch force field~\cite{buch1994path} was applied to model hydrogen molecules as gas adsorbates. Simulations were conducted as follows. First, $2\times2\times2$ supercell was equilibrated under consecutive NPT, NVT, and NVE ensembles for 2 ns, 2 ns, and 1 ns, respectively. Then, heat flux data were collected in subsequent NVE ensemble for 1 ns to compute the thermal conductivities using Green-Kubo formalism~\cite{chen2005gk}, with a total correlation time of 8 ps (see \SM~Sec.~I for details). Five independent MD simulations were performed to obtain averaged thermal conductivity of each system. All simulations were run with a timestep of 1 fs. 

We also conducted non-equilibrium MD (NEMD) simulations to obtain spectral thermal conductivities and the accumulations~\cite{saaskilahti2014role,saaskilahti2016spectral,saaskilahti2015frequency}. We constructed $2\times2\times6$ supercell, consisting of 13248 atoms and performed the NEMD simulations with LAMMPS with a timestep of 0.5 fs using the following procedures. First, the domain was equilibrated in NPT ensemble for 1.5 ns. Next, two fixed regions were placed at both ends of the cell, which was relaxed in NVT ensemble for an additional 1.5 ns. During the subsequent NVE ensemble for 1 ns, we added two Langevin thermostats adjacent to the fixed regions, to set a temperature difference of 80 K and to generate a heat flux along the z-direction. Then, an initial 2.5 ns main run was carried out to calculate the temperature gradient in NVE ensemble, which was determined using the best linear fit of the steady-state temperature profile (see \SM~Sec.~II for a representative profile). Frequency-dependent heat current was obtained for an additional 375 ps in NVE ensemble, which can be expressed as

\begin{equation}
Q(\omega) = \frac{2}{t_s} \, \mathrm{Re} \sum_{i \in left} \sum_{j \in right} \left\langle \mathbf{F}_{ij}(\omega) \cdot \mathbf{v}_i(\omega)^* \right\rangle
\label{eq:specheat}
\end{equation}

where $t_s$ is the simulation time and $\mathbf{F}_{ij}$ is the interatomic force acting on atom with index $i$ (located left to middle of simulation domain) due to atom $j$ (located right to middle of simulation domain). Then, we computed the thermal conductivity accumulation function given by
\begin{equation}
\kappa(\omega) = \int_{0}^{\omega}\frac{Q(\omega')}{A\cdot\nabla T}d\omega'
\label{eq:accumkap}
\end{equation}

where $A$ is cross-sectional area and $\nabla T$ is temperature gradient. 

\section{Results and discussions}
    \begin{figure}
{\includegraphics[width=12cm,keepaspectratio]{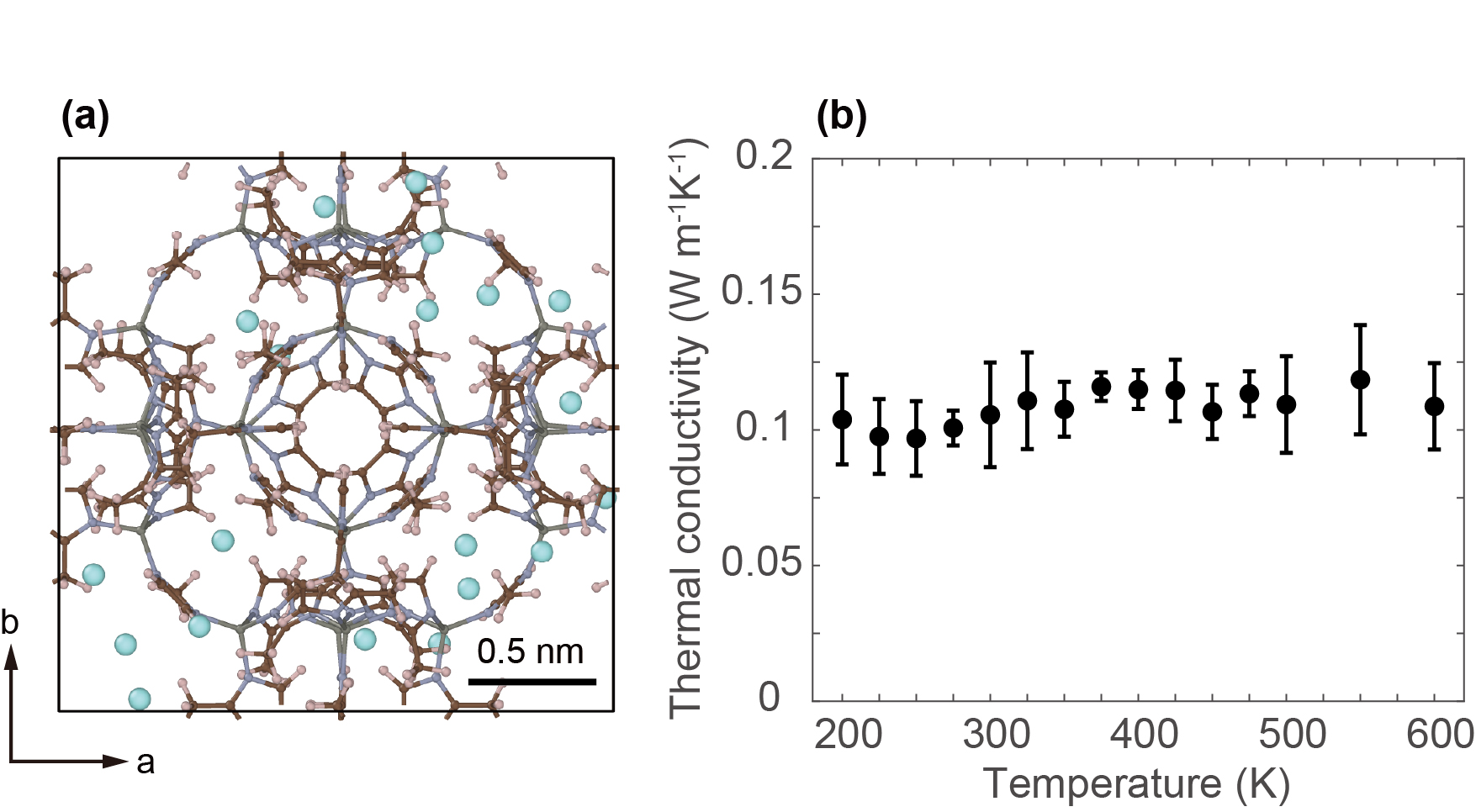}}
    \caption{\label{fig:fig1} 
(a) Schematic of the molecular structures of ZIF-lta infiltrated with hydrogen molecules (H$_2$) at a gas density of 1.75 molecules nm$^{-3}$ (zinc: silver colors; carbon: brown colors; nitrogen: blue colors; hydrogen: pink colors; hydrogen molecules: cyan colors). Note that hydrogen molecules are modeled as point particles using single-site spherical potentials based on Buch force field.
(b) Thermal conductivity versus temperature ranging from 200 K to 600 K. Over the temperatures considered here, the thermal conductivity shows weak temperature dependence.
}
\end{figure}

We calculated bulk thermal conductivities at a range of temperatures for pristine ZIF-lta using EMD. The unit cell of ZIF-lta is shown in Fig.~\ref{fig:fig1}(a), which consists of 552 atoms with a lattice constant of 22.51 \AA; each Zn atom is tetrahedrally bonded to four adjacent N atoms in 2-methylimidazolate organic linkers.  We note that classical MD simulation tends to overestimate heat capacity and the resulting thermal conductivity, especially at temperatures below the Debye temperature~\cite{huang2007MD,zhou2018classic}. To avoid such issues, the minimum temperature considered in this study is set to 200 K, which exceeds the estimated Debye temperature (see \SM~Sec.~III for details). As shown in Fig.~\ref{fig:fig1}(b), pristine ZIF-lta has room temperature bulk thermal conductivity of $\sim 0.10$~\wmk, in a quantitative agreement with a prior prediction~\cite{cheng2021correlation}. We note that the thermal conductivity does not show a marked temperature dependence.

    \begin{figure}
 {\includegraphics[width=13cm, keepaspectratio]{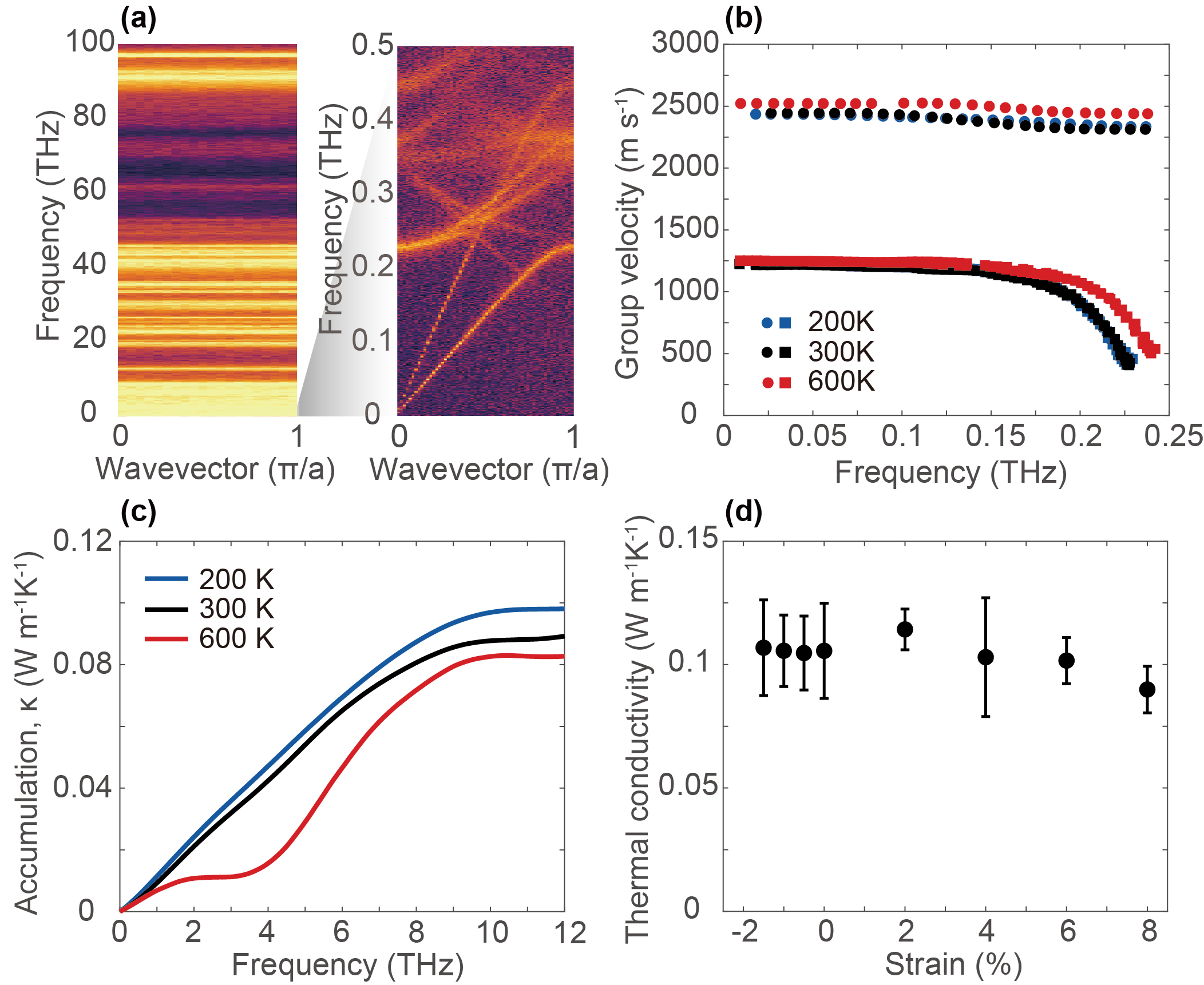}}
\caption{\label{fig:fig2}
(a) Calculated SED showing phonon dispersion with branches up to 100 THz (left panel). Also shown is the dispersion below 0.5 THz (right panel), in which overlapping between acoustic polarizations and many other broadened modes is clearly visible. In both panels, $a$ denotes the lattice constant.
(b) Extracted group velocities versus frequency for acoustic modes. Group velocities were extracted as the slope of the  dispersion curves in (a), yielded by peak positions of best Lorentzian fits (LA modes: circles; TA modes: squares). As the temperature decreases, group velocities decrease, indicating the occurrence of phonon softening.
(c) Thermal conductivity accumulation computed using NEMD. The majority of accumulated thermal conductivity arises from vibrational modes above 0.5 THz.
(d) Thermal conductivity versus hydrostatic strain. At tensile strain with magnitude above 2\%, thermal conductivity marginally decreases.
}
\end{figure}

Following Refs.~\cite{thomas2014SED,sungjae2025cof}, we extracted phonon dispersion and phonon transport properties of pristine ZIF-lta using spectral energy density (SED). Figure~\ref{fig:fig2}(a) shows SED representing phonon dispersion of ZIF-lta across the full frequency range along with that within the sub-THz region. Full dispersion shows broadened bands extending up to 100 THz due to the presence of light C and H atoms, as confirmed by our vibrational density of states (VDOS) spectra (see \SM~Sec.~IV for additional data), which is also observed in previous studies \cite{wang2015anisotropic}. In a zoomed-in view of acoustic region, despite the overall broadening visible in many other modes, we observe well-defined phonon modes near $\Gamma$ point. However, as the wavevector approaches the Brillouin zone edge, acoustic phonon branches become increasingly broadened, indicating significantly reduced MFPs of phonon modes that are approaching the Ioffe-Regel limit. More precisely, we observed that the modes with wavevector above 0.8~${\pi/a}$ exhibit MFPs comparable to the wavelengths, in analogy to diffusons, which are non-propagating yet delocalized modes in amorphous solids (see \SM~Sec.~V for details). The results indicate a high degree of anharmonicity, despite long-range crystalline order in ZIF-lta, as also reported in other crystalline solids such as higher manganese silicides~\cite{chen2015mn} and tin selenide~\cite{kang2019ts}. These findings further imply that modes with frequencies above the acoustic range are highly likely to exhibit distinctly short MFPs, as indicated by substantially broadened bands shown in Fig.~\ref{fig:fig2}(a).

Figure.~\ref{fig:fig2}(b) shows calculated acoustic group velocities, which are $\sim$ 2400 \velocity~for LA branch and $\sim$ 1200 \velocity~for TA branches. These values are in a quantitative agreement with those estimated using adiabatic compressibility (see \SM~Sec.~III). As temperature decreases from 600 K to 200 K, the group velocities near $\Gamma$ point decrease by 4\% for the LA modes and by 3\% for the TA modes. This reduction in group velocities at lower temperatures indicates phonon softening, in contrast to conventional crystalline solids, such as silicon~\cite{kuryliuk2019silicon} and boron arsenide~\cite{kang2019boron}, where the acoustic mode group velocities marginally increase at lower temperatures.

We additionally calculated the Gruneisen parameter to characterize the anharmonicity. Following identical procedure in Ref.~\cite{robbins2015poly}, we extracted Gruneisen parameter defined as $\gamma = - {\partial\mathrm{ln}\omega}/{\partial\mathrm{ln}V}$, in which $\omega$ and $V$ correspond to LA phonon frequency and volume, respectively, which change under applied strain. Averaging across the LA mode yields $\gamma=-1.36$, which is considerably larger than that reported in MOF-5~\cite{huang2007MD}. This relatively high Gruneisen parameter indicates strong anharmonicity, which is attributed to a more complex unit cell of ZIF-lta, containing a greater number of atoms (ZIF-lta: 552 atoms vs. MOF-5: 424 atoms) and metal centers (ZIF-lta: 24 Zn atoms vs. MOF-5: 4 Zn atoms). Such pronounced anharmonicity strongly suppresses the lifetimes of many phonon branches, consistent with the broadened peaks observed in Fig.~\ref{fig:fig2}(a), resulting in exceptionally low thermal conductivity. According to Ref.~\cite{wang2015anisotropic}, this anharmonicity primarily originates from strong anharmonic bonding between metal nodes and organic linkers. As a result, phonons mainly scatter at metal sites, as evidenced by significantly high Gruneisen parameter of metal atoms compared to other species~\cite{wang2015anisotropic}. We further note that the negative sign in $\gamma$ indicates that bond compression reduces phonon frequencies, as observed in MOFs such as HKUST-1~\cite{fan2024anomalous}, but not visible in typical crystals like silicon~\cite{kuryliuk2019silicon}.

We also calculated spectral thermal conductivity along with the thermal conductivity accumulation function by performing NEMD calculations. Details are provided in \nameref{sec:method}, and the resulting calculations yielded by Eqs.~\ref{eq:specheat} and~\ref{eq:accumkap} at several temperatures are presented in Fig.~\ref{fig:fig2}(c). As a validation of our NEMD calculations, we compared the maximum of spectral thermal conductivity accumulation with our EMD results in Fig.~\ref{fig:fig1}(b), which we observed a quantitative agreement between them. We also observe that dominant contribution to the thermal conductivity comes from the modes above 0.5 THz. As previously discussed, since the modes above the corresponding frequency are expected to exhibit markedly short MFPs, as observed by significantly broadened bands compared to acoustic modes in Fig.~\ref{fig:fig1}(a), our results suggest that heat is mainly transported by strongly diffusive modes that satisfy Ioffe-Regel criteria. As a result, overall thermal conductivity is exceedingly small, while showing lacked temperature dependence. Our findings are consistent with prior reports on MOF-5~\cite{huang2007MD} and crystalline porous materials such as zeolite-A~\cite{mcgaughey2004porous}, which exhibit unusual temperature dependence of the thermal conductivity despite long-range order, attributed to modes with MFPs on the order of lattice parameter.

We next examined hydrostatic strain dependent thermal conductivity of ZIF-lta, as shown in Fig.~\ref{fig:fig2}(d), in which we applied compressive strains only up to $\sim 2$\% to preserve crystallinity~\cite{weng2019zifff} (see \SM~Sec.~VI for discussions). Within strains magnitude below 2\%, thermal conductivity does not show a clear dependence on the applied strain. In contrast, a decrease in thermal conductivity on the order of $\sim20$\% can be seen under 8\% tensile strain, albeit with relatively large error bars. We attribute the observed reduction in the thermal conductivity under tensile strain to a decrease in volumetric heat capacity, which scales inversely with volume in MD simulations. However, we note that for small strains below 2\%, this scaling is marginal to yield an apparent change in the thermal conductivity. We additionally note that a prior work reported on HKUST-1 has shown that strain-dependent thermal conductivity can be affected by phonon group velocities and lifetimes of low-frequency phonons below 2 THz, rather than by heat capacity~\cite{fan2024anomalous}. In our case, despite observed increases in MFPs and group velocities of phonons below 0.3 THz under tensile strain (see \SM~Sec.~VII for details), such strain-induced changes have limited impact on thermal conductivity relative to that of heat capacity due to suppressed contribution of low-frequency vibrational modes below 1 THz in ZIF-lta.

\begin{figure}
{\includegraphics[width=13cm, keepaspectratio]{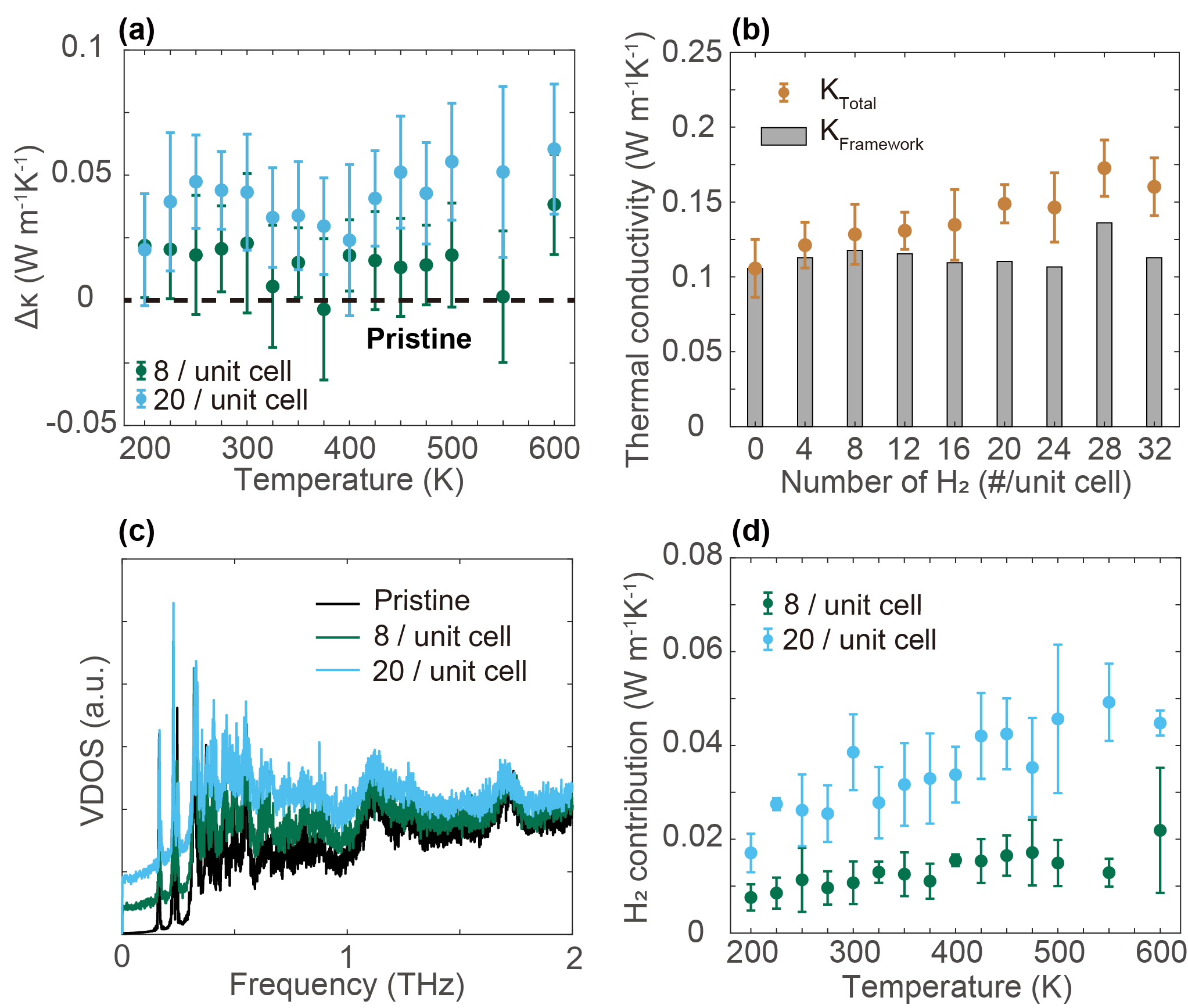}}
\caption{\label{fig:fig3}
(a) Relative change in thermal conductivity of hydrogen infiltrated structure, compared to pristine ZIF-lta. In general, infiltrated structures exhibit higher thermal conductivities than pristine case.
(b) Thermal conductivity versus the number of hydrogen molecules per unit cell (orange circles: total thermal conductivity contributed by both adsorbates and solid frameworks; gray bars: thermal conductivity contributed by solid frameworks only). Noting that the solid contributions are larger compared to that of pristine, total thermal conductivity monotonically increases as the number of hydrogen molecules increases.
(c) Calculated VDOS for frequency below 2 THz. Compared to pristine case, the VDOS intensity of the ZIF-lta with infiltrated hydrogen molecules is clearly larger, indicating that additional vibrational modes are present.
(d) Thermal conductivity contributed by infiltrated hydrogen molecules versus temperature. For the temperature range considered here, thermal conductivity is observed to be increased by factor of $\sim 2$ for both cases. In all cases, green color represents adsorption of 8 hydrogen molecules per unit cell (corresponding density: 0.7 molecules nm$^{-3}$), and blue color represents adsorption of 20 hydrogen molecules per unit cell (corresponding density: 1.75 molecules nm$^{-3}$).
}
\end{figure}

We next provide an insight into the impact of adsorbates on the thermal transport in ZIF-lta by calculating thermal conductivity of hydrogen infiltrated ZIF-lta. The difference in the thermal conductivity between infiltrated structure and the pristine case is shown in Fig.~\ref{fig:fig3}(a). The guest-infiltrated structure shows enhanced thermal conductivity compared to pristine case in given temperature range, which becomes more pronounced as the number of adsorbates increases. We further note that infiltrated structures show relatively stronger temperature dependence, compared to pristine case as shown in Fig.~\ref{fig:fig1}(b).

We calculated total thermal conductivity and the framework contribution upon gas adsorption as a function of the number of hydrogen molecules. The framework contribution was obtained by computing the cross-correlation between the heat flux of the framework and that of the total system, including both the framework and the gas molecules. As in Fig.~\ref{fig:fig3}(b), we see that total apparent thermal conductivity increases with the number of hydrogen molecules, suggesting that gas adsorption leads to enhanced system's thermal conductivity. We note that, despite the uncertainty, the thermal conductivity contributed by the framework is larger compared to pristine case. According to Ref.~\cite{babaei2016ideal}, the only mechanism to decrease framework contribution upon gas adsorption is stronger anharmonic scattering, while increased framework contribution has been attributed to improved thermal boundary conductance between gas and framework~\cite{giri2021heat}. Our results suggest that the most likely cause of enhanced heat transport is latter effect that outweighs the former, leading to an increase in framework thermal conductivity contribution, which becomes more pronounced at higher adsorbate density.

We then calculated the VDOS for pristine and hydrogen infiltrated ZIF-lta. Figure~\ref{fig:fig3}(c) presents VDOS spectra below 2 THz before and after adsorption. While the full VDOS spectra is difficult to discern (see \SM~Sec.~IV for additional data), hydrogen gas adsorption leads to an increase of VDOS in the given frequency range, indicating additional vibrational modes contribution from hydrogen molecules. We highlight that additional low-frequency vibrational modes upon gas adsorption are formed by gas molecules, despite the thermal transport in pristine case being dominated by short-range modes that are close to Ioffe-Regel limit. According to Ref.~\cite{thakur2025co2}, the vibrational spectra of adsorbates shifts to higher frequency compared to bulk gas, as gas molecules collide more frequently with themselves and with the framework. Consistent with the previous report, we observe the VDOS of hydrogen molecules is red-shifted, as a result of frequent gas-gas and gas-framework interactions (see \SM~Sec.~VIII for details).

We next investigated the impact of temperature on the thermal conductivity contributed by hydrogen molecules. Results are presented in Fig.~\ref{fig:fig3}(d). We observed that contribution of hydrogen molecules to thermal conductivity monotonically increases by a factor of $\sim2$ between 200 K and 600 K for both cases. We also note that mass diffusivity of infiltrated hydrogen molecules increases with temperature due to enhanced molecular mobility (see \SM~Sec.~IX for details), which aligns with the trend observed in hydrogen contribution to thermal conductivity. Furthermore, according to Refs.~\cite{yamaguchi2024aniso, giri2021heat}, suppressed diffusivity of adsorbates in the framework compared to bulk case indicates that thermal transport is governed by conduction. We observed that diffusivity of hydrogen molecules in ZIF-lta ($\sim0.001$ cm$^2$ s$^{-1}$) is three orders of magnitude lower than that of bulk gas (1.4 cm$^2$ s$^{-1}$~\cite{yang2005self}). This suppressed diffusivity indicates that the gas mobility is significantly reduced by spatial confinement of gas molecules within the framework, further implying that heat is transported by conduction, rather than by convection as supported by previous works~\cite{yamaguchi2024aniso, giri2021heat}.

Combining all these results, the enhanced thermal conductivity upon hydrogen gas adsorption in ZIF-lta, as shown in Fig.~\ref{fig:fig3}(a), is primarily attributed to the efficient energy transport between gas molecules and enhanced thermal conductance between gas and framework, which becomes more pronounced at higher gas densities, as evidenced by VDOS that shows high intensities, particularly at low frequencies. 
Particularly, we observed that thermal conductivity of hydrogen molecules confined in the framework is higher than the conductive contribution of bulk gas at even higher gas density (see \SM~Sec.~X for details). As our diffusivity calculations show heat is mainly carried via conduction rather than through mass transport of gas molecules in ZIF-lta, these results indicate that gas-framework interactions enhance the apparent thermal conductivity. Additionally, marked temperature dependence of total thermal conductivity in infiltrated structure, relative to pristine case, is mainly attributed to higher kinetic energy of gas molecules at elevated temperatures. This, in turn, induces more frequent gas-gas interactions, while enhancing thermal conductance between gas and framework.

We discuss two primary conclusions in our findings. First, ZIF-lta exhibits pronounced anharmonicity, as is also observed in organic polymers such as polynorbornene (PNb)~\cite{robbins2015poly}, which suppresses the formation of long-lived phonons. As a result, thermal transport is mainly dominated by vibrational modes that satisfy Ioffe-Regel criteria, similar to diffusons in amorphous materials, consistent with prior reports on ZIF-4 and ZIF-62~\cite{zhou2022origin}. However, unlike PNb, we attribute origin of anharmonicity emerging in ZIF-lta to strong anharmonic scattering occurring at metal sites, which are absent in PNb. 

Second, we find that hydrogen adsorption in ZIF-lta increases overall thermal conductivity, as enhanced heat transfer facilitated by kinetic energy of adsorbates surpasses the effect of additional anharmonic scattering at the framework, which impedes the heat transport. In HKUST-1, it has been reported that adsorbed hydrogen molecules behave mainly like phonon scatterers, leading to a reduction in thermal conductivity by up to $\sim50$\%~\cite{fan2025influence}. On the other hand, both hydrogen and deuterium adsorption in MOF-5 results in enhanced overall thermal conductivity, with the effect more pronounced for deuterium molecules~\cite{han2014relation}. We note that the observed enhancement in~Ref.~\cite{han2014relation} for both cases are likely to be attributed to enhanced thermal conductance between lattice and gas, rather than convection, since deuterium molecules, having a higher mass and thus lower diffusivity, would be expected to contribute less via their mass transport. Our results are consistent with Ref.~\cite{han2014relation} in that hydrogen gas in ZIF-lta enhances overall thermal conductivity, mainly through conduction by both gas-gas and gas-framework interactions.

\section{Conclusions}
In summary, we report molecular dynamics investigations of thermal transport in ZIF-lta. Despite long-range crystalline order, ZIF-lta shows ultra-low thermal conductivity lacking temperature dependence. We show that heat transport in ZIF-lta is governed by thermal phonons with mean free paths comparable to their wavelength, analogous to diffusons in amorphous materials owing to strong anharmonicity caused by complex unit cell with large number of metal centers. We further observe that hydrogen adsorption increases in total apparent thermal conductivity in them, mainly contributed by additional vibrational modes due to gas-gas and gas-framework interactions. Our work provides comprehensive insights into heat transport in MOFs and proposes means to modify their heat conduction by gas infiltrations.

\section*{Supporting information}
Calculated thermal conductivities versus correlation time computed using Green-Kubo formlism, temperature profile using Non-equilibrium molecular dynamics (NEMD) simulations, estimation of Debye temperature, additional data of vibrations density of states (VDOS), spectral mean free paths (MFPs) for longitudinal acoustic (LA) polarization, calculation of pressure dependent density, acoustic phonon properties in ZIF-lta with applied strain, VDOS comparision to bulk hydrogen gas, calculation of hydrogen diffusivity, and comparative analysis of the thermal conductivity of bulk hydrogen gas and that of infiltrated hydrogen in ZIF-lta.\\

\section*{Acknowledgments}
 This work was supported by the National Research Foundation of Korea (NRF) grant funded by the Korea government(MSIT)(No. RS-2023-00211070) and the National Supercomputing Center with supercomputing resources including technical support (KSC-2023-CRE-0537).

\section*{Author contributions}
T.K. conceived and supervised the project. H.O. performed the MD calculation and analyzed the data. H.O. and T.K. wrote the manuscript.
 
\section*{Data availability}
The authors declare that the data supporting the findings of this study are available within the paper and its Supporting Information files.
 
\section*{Competing interests.}
The authors declare no competing interests.

\bibliographystyle{unsrtnat}

\bibliography{MOFRef}

\clearpage
\begin{figure}
\centering
\includegraphics[width=8.25cm,keepaspectratio]{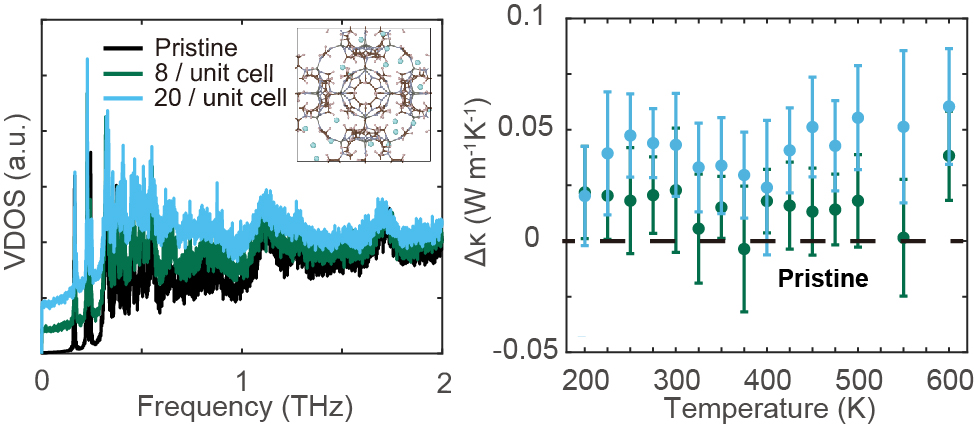}
\end{figure}
For Table of Contents Only.

\end{document}


\title{Supporting information: Thermal transport and the impact of hydrogen adsorption in Linde Type A zeolitic imidazolate frameworks}
\author{Hyunseok Oh~\orcidicon{0009-0001-2571-8209}}
\affiliation{%
Department of Mechanical Engineering, Seoul National University, Seoul 08826, Republic of Korea}%
\author{Taeyong Kim~\orcidicon{0000-0003-2452-1065} }
 \email{tkkim@snu.ac.kr}
\affiliation{%
Department of Mechanical Engineering, Seoul National University, Seoul 08826, Republic of Korea}%
\affiliation{
Institute of Advanced Machines and Design, Seoul National University, Seoul 08826, Republic of Korea
}
\affiliation{%
Inter-University Semiconductor Research Center, Seoul National University, Seoul 08826, Republic of Korea}%

\date{\today}
{    \global\let\newpagegood\newpage
    \global\let\newpage\relax
\maketitle}

\clearpage

\section{Calculated thermal conductivities versus correlation time computed using Green-Kubo formlism}
\begin{figure}[hbt!] 
\includegraphics[width=\textwidth,height=.35\textheight,keepaspectratio]{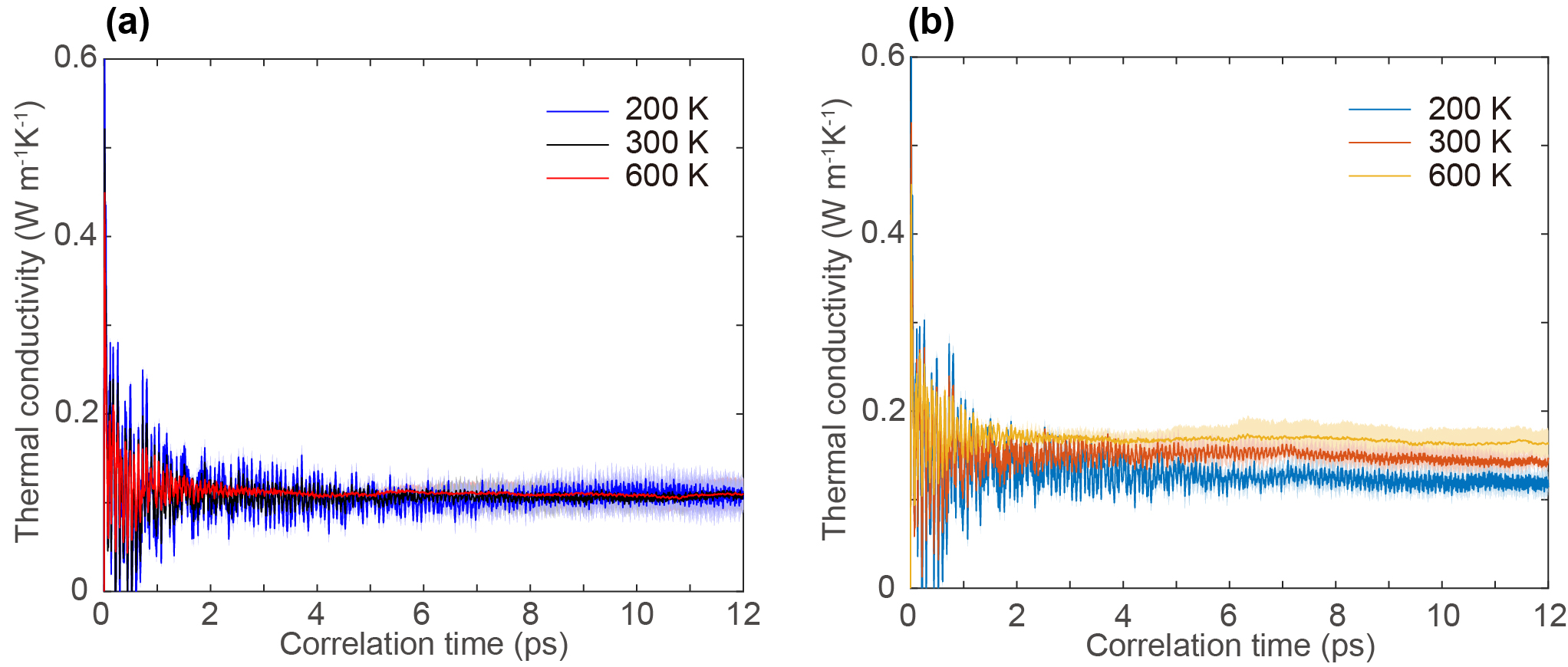}
\caption{Representative computed running thermal conductivities using the Green-Kubo formalism at several temperatures for (a) pristine structure and (b) hydrogen infiltrated structure (1.75 molecules nm$^{-3}$). Solid lines correspond to average values, while shaded areas correspond to standard deviation determined from five independent runs.}
\label{SIfig:green_kubo}
\end{figure}

In this section, we provide results of calculated running thermal conductivities of ZIF-lta computed using Green-Kubo formalism. The thermal conductivities versus correlation time at several temperatures are given in Fig.~\ref{SIfig:green_kubo}. The thermal conductivities generally exhibit initial transient features by approximately $\sim 2$ ps, after which the values are stabilized. Based on these observations, we chose a correlation time of 8 ps, which is sufficiently long enough to ensure the convergence of the thermal conductivity to a constant value, while reducing the fluctuations. We note that, in determining the thermal conductivity for each run, we time-averaged the values over a 0.2 ps centered around the selected correlation time. The standard deviation was obtained from five independent runs.

\clearpage

\section{Steady-state temperature profile in Non-equilibrium molecular dynamics (NEMD) simulations}
\begin{figure}[hbt!] 
\includegraphics[width=\textwidth,height=.35\textheight,keepaspectratio]{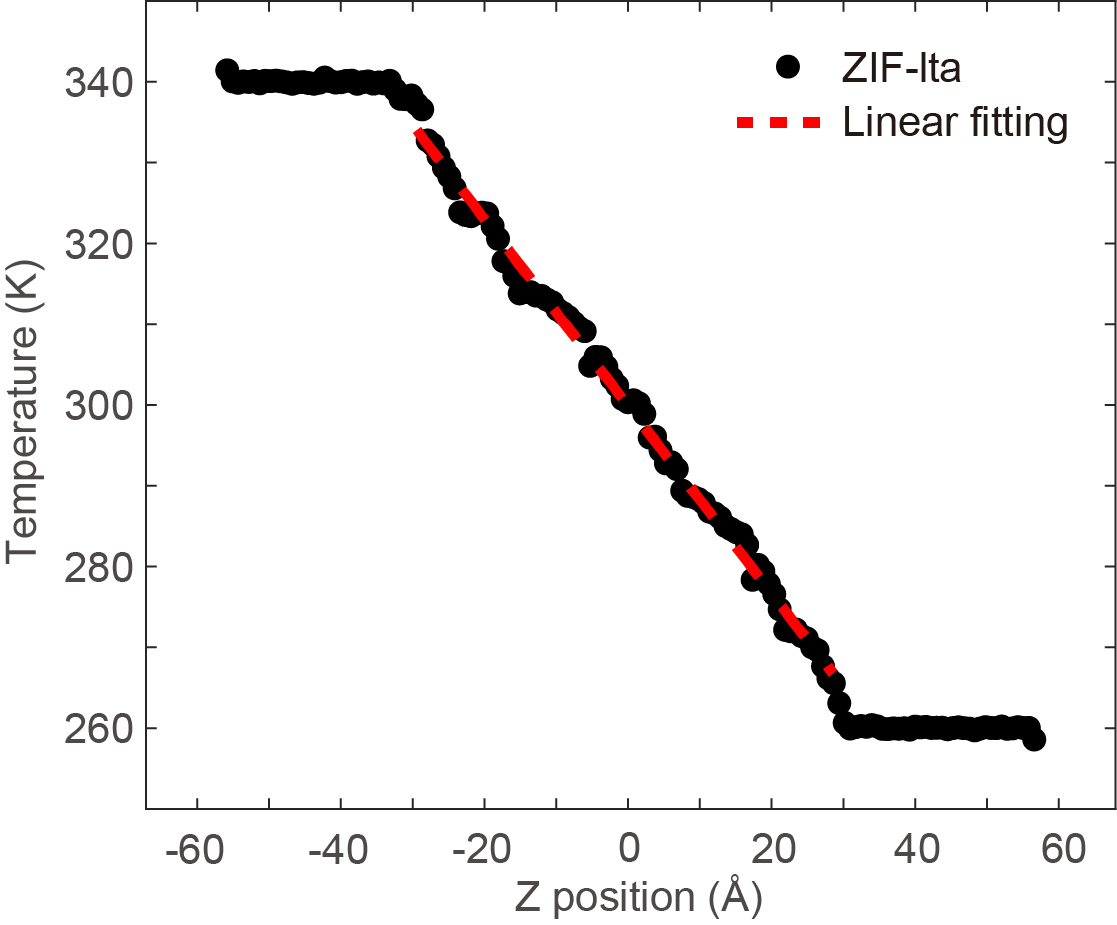}
\caption{Representative temperature distribution of ZIF-lta at 300 K. The red dashed line is the best linear fit of the steady-state temperature profile.}
\label{SIfig:temperature_profile}
\end{figure}

In this section, we provide a representative calculated steady-state temperature profile along with the best linear fit at 300 K in our NEMD simulations. As shown in Fig.~\ref{SIfig:temperature_profile}, we observe kinks at several z positions, which correspond to junctions between metal nodes and organic linkers. According to Ref.~\cite{Wieser2021temp}, the kinks indicate that heat transport is primarily impeded at the junctions between metal nodes and organic linkers.

\clearpage

\section{Estimation of Debye temperature}
\begin{figure}[hbt!] 
\includegraphics[width=\textwidth,height=.35\textheight,keepaspectratio]{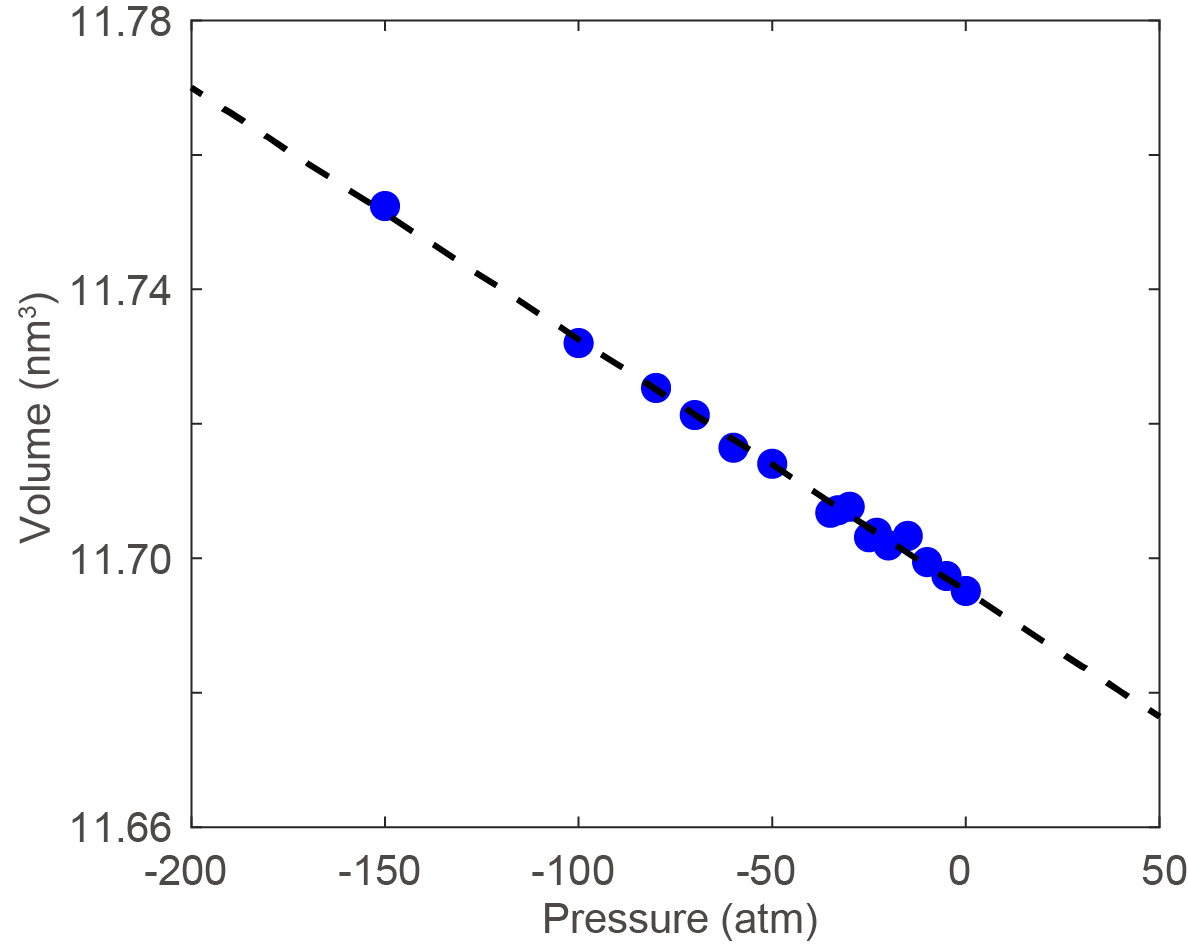}
\caption{Calculated pressure-volume relations of ZIF-lta under adiabatic conditions. Blue circles correspond to values yielded by MD simulations, while black-dashed line corresponds to best linear fit.}
\label{SIfig:volpress}
\end{figure}

In this section, we describe our estimation of the Debye temperature of ZIF-lta. The Debye temperature can be estimated using 

\begin{equation}
T_D=\frac{\hbar}{k_b}v_g(6\pi^2\rho_N)^{1/3}
\end{equation}

where $\hbar$ is the reduced Planck constant, $k_b$ is the Boltzmann constant, $v_g$ is group velocity, and $\rho_N$ is atomic number density~\cite{kittel2018introduction}. Noting that ZIF-lta is thermally isotropic, the average group velocity was estimated using longitudinal group velocity ($v_{g,L}$) and transverse group velocity ($v_{g,T}$) from
\begin{equation}
v_{g,L}=[\frac{3(1-2\nu)}{k_s\rho}]^{1/2}
\label{Eq:vL}
\end{equation}
\begin{equation}
v_{g,T}=[\frac{3(1-2\nu)}{2(1+\nu)k_s\rho}]^{1/2}
\label{Eq:vT}
\end{equation}
where $\rho$ is density, $k_s$ is adiabatic compressibility, and $\nu$ is the Poisson's ratio. We took the Poisson's ratio to be 0.3~\cite{huang2007md}.
Following the identical procedure in Ref.~\cite{zhou2022origin}, we estimated the adiabatic compressibility using
\begin{equation}
k_s=-\frac{1}{V}(\frac{\partial V}{\partial P})
\label{Eq:ks}
\end{equation}
where $P$ is pressure and $V$ is volume. ${\partial V}/{\partial P}$ was extracted using the best linear fit that are compatible with isoenthalpic-isobaric simulation given in Fig.~\ref{SIfig:volpress}. Resulting calculations using Eqs.~\ref{Eq:vL},~\ref{Eq:vT}, and~\ref{Eq:ks}~yield $k_s\sim3.2\times10^{-10}$ Pa$^{-1}$, $v_{g,L} \sim2200$m s$^{-1}$ and $v_{g,T}\sim1400$m s$^{-1}$, respectively. Finally, Debye temperature was estimated to be $\sim162$ K.

\clearpage

\section{Additional data of vibrational density of states (VDOS)}
This section presents VDOS of ZIF-lta. Briefly, we calculated the VDOS by taking the Fourier transform of atomic velocity autocorrelation function (VACF)~that can be expressed as~\cite{dickey1969computer}.
\begin{equation}
VDOS(\omega) \propto \int_{-\infty}^{\infty} VACF(t) \, e^{-i 2 \pi \omega t} \, dt
\label{eq:VDOSexp}
\end{equation}
In Eq.~\ref{eq:VDOSexp}, VACF is defined as $VACF(t) = \langle \sum_i \mathbf{v}_i(0) \cdot \mathbf{v}_i(t)\rangle/\langle \sum_i \mathbf{v}_i(0) \cdot \mathbf{v}_i(0) \rangle $.

\begin{figure}[hbt!] 
\includegraphics[width=\textwidth,height=.5\textheight,keepaspectratio]{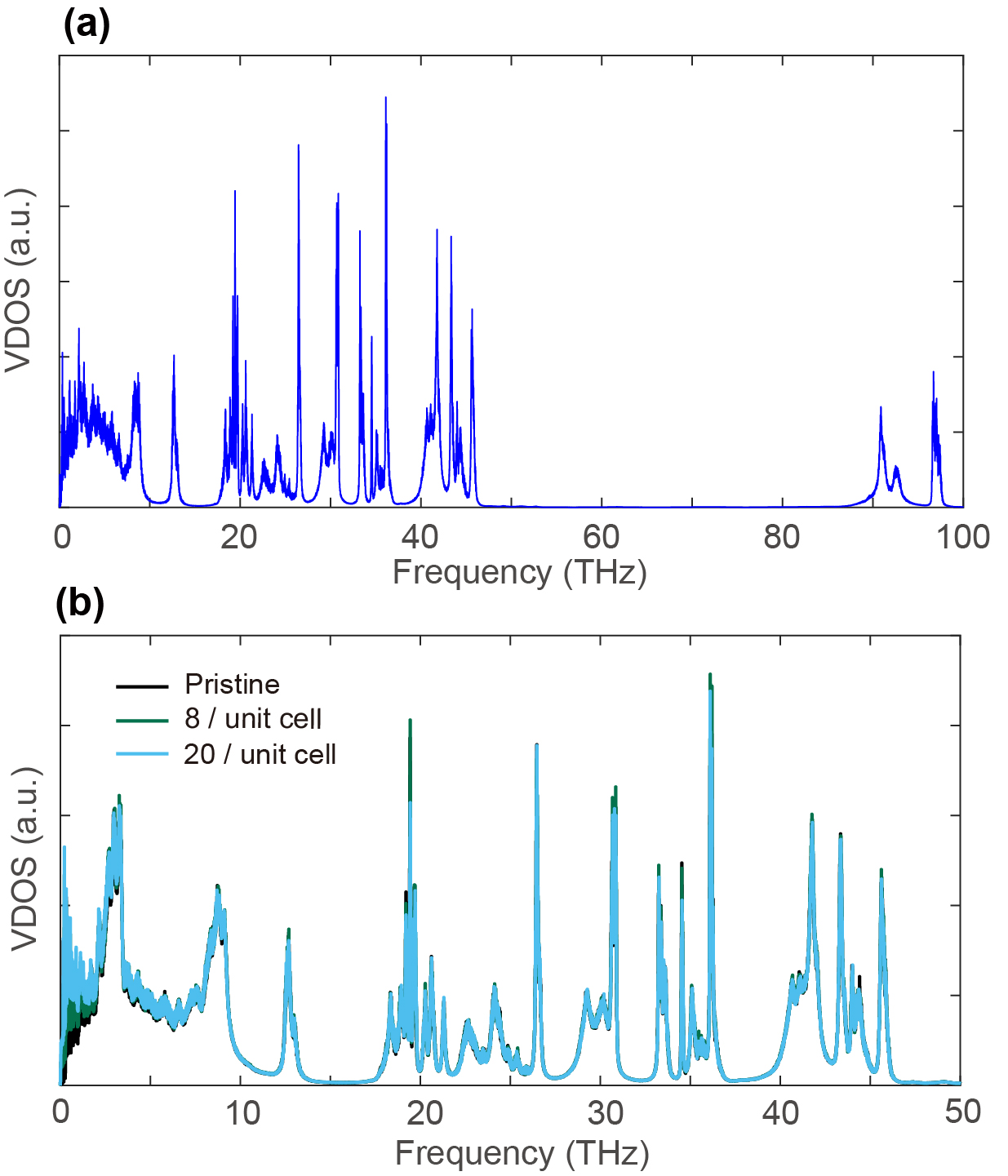}
\caption{(a) Calculated VDOS of ZIF-lta below 100 THz at room temperature. Distinct peaks are visible up to 100 THz. (b) Calculated VDOS for frequency below 50 THz (pristine: black line; ZIF-
lta with 8 hydrogen gas molecules per unit cell: green line; that with 20 hydrogen gas molecules per unit cell: blue line)}
\label{SIfig:VDOS}
\end{figure}

Resulting calculations using Eqs.~\ref{eq:VDOSexp} in pristine case are presented in Fig.~\ref{SIfig:VDOS}(a), which shows distinct peaks up to 100 THz at room temperature. The results for ZIF-lta before/after hydrogen molecules adsorption below 50 THz are shown in Fig.~\ref{SIfig:VDOS}(b). We note that the VDOS is difficult to discern except frequencies below $\sim2$ THz, which are given in the main text.

\clearpage

\section{Spectral Mean free paths (MFPs) for longitudinal acoustic (LA) polarization}
\begin{figure}[hbt!] 
\includegraphics[width=\textwidth,height=.3\textheight,keepaspectratio]{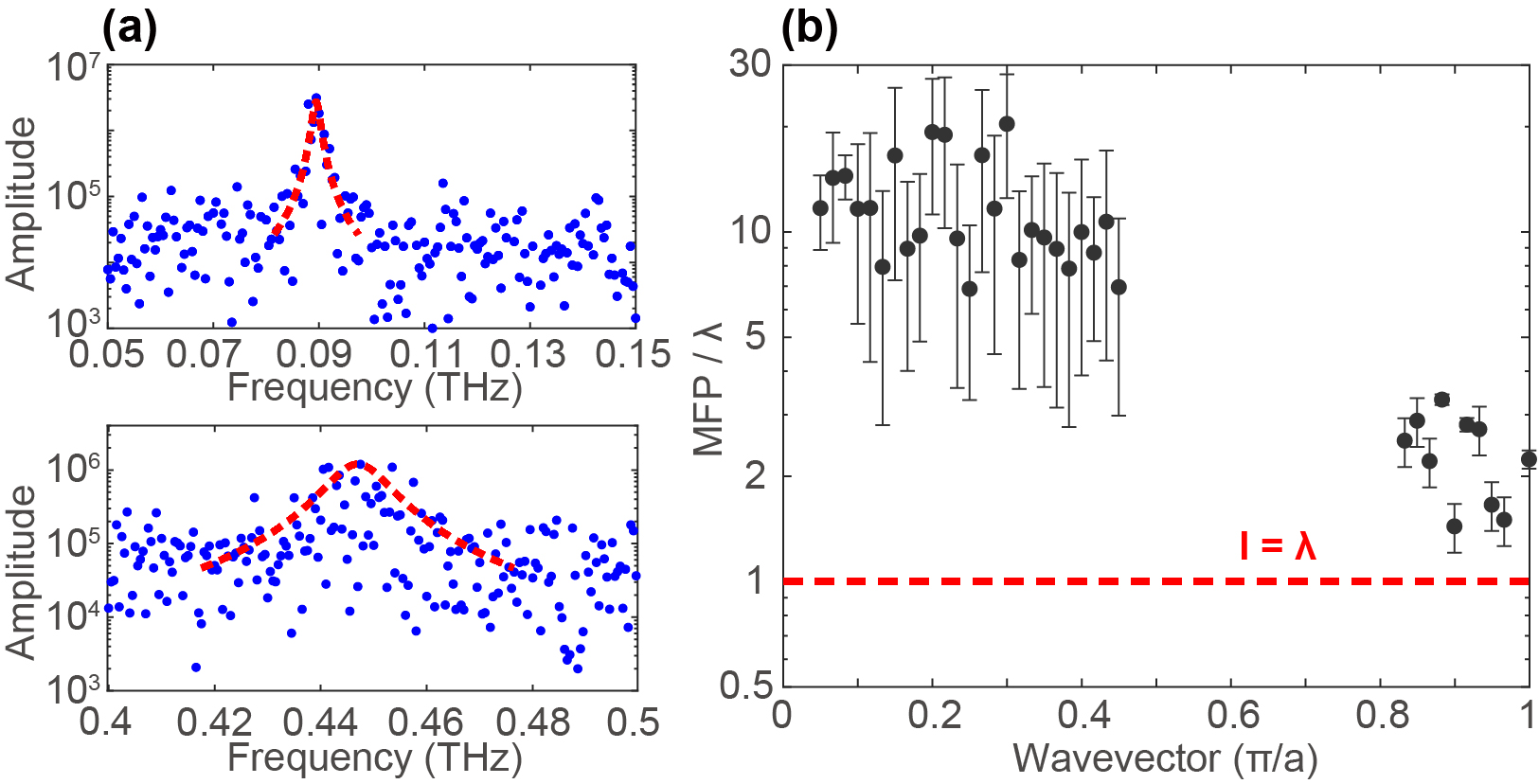}
\caption{Vertical line-cut of SED for LA polarization (blue dots) and corresponding best Lorentzian fit (red dashed line) for frequency near zone center (top) and for those near zone edge (bottom). (b) Spectral mean free paths normalized by wavelength versus wavevector for LA modes. ($a$ denotes the lattice constant of 22.51 \AA). Red dashed line indicates the Ioffe-Regel crossover. As the wavevector increases, $l$ approaches $\lambda$, satisfying the condition for Ioffe-Regel crossover.
}
\label{SIfig:SEDfit}
\end{figure}

In this section, we present our analysis of spectral MFPs for LA modes. Figure~\ref{SIfig:SEDfit}(a) shows vertical line-cut of spectral energy density (SED) for LA mode along with the corresponding best Lorentzian fit that yields phonon lifetime ($\tau$). Noting that inverse of the full-width at half-maximum (line-width) of the Lorentzian fit is phonon lifetime, we observe that modes near Brillouin zone edge clearly exhibit substantially smaller $\tau$, relative to those near the zone center.

Spectral MFPs ($l$) are given by $l = v_{g} \tau$. Figure~\ref{SIfig:SEDfit}(b) shows $l$ normalized by wavelength ($\lambda$) versus the wavevector. As the wavevector increases, $l$ approaches the $\lambda$, satisfying the condition for Ioffe-Regel crossover, further implying that modes above acoustic frequency range are expected to show MFPs comparable to their wavelengths. At wavevectors ranging from 0.4 ${\pi/a}$ to 0.8 ${\pi/a}$, obtaining $l$ was nearly impossible due to the overlap with substantially broadened optical modes.

\clearpage

\section{Calculation of pressure dependent density}
\begin{figure}[hbt!] 
\includegraphics[width=\textwidth,height=.35\textheight,keepaspectratio]{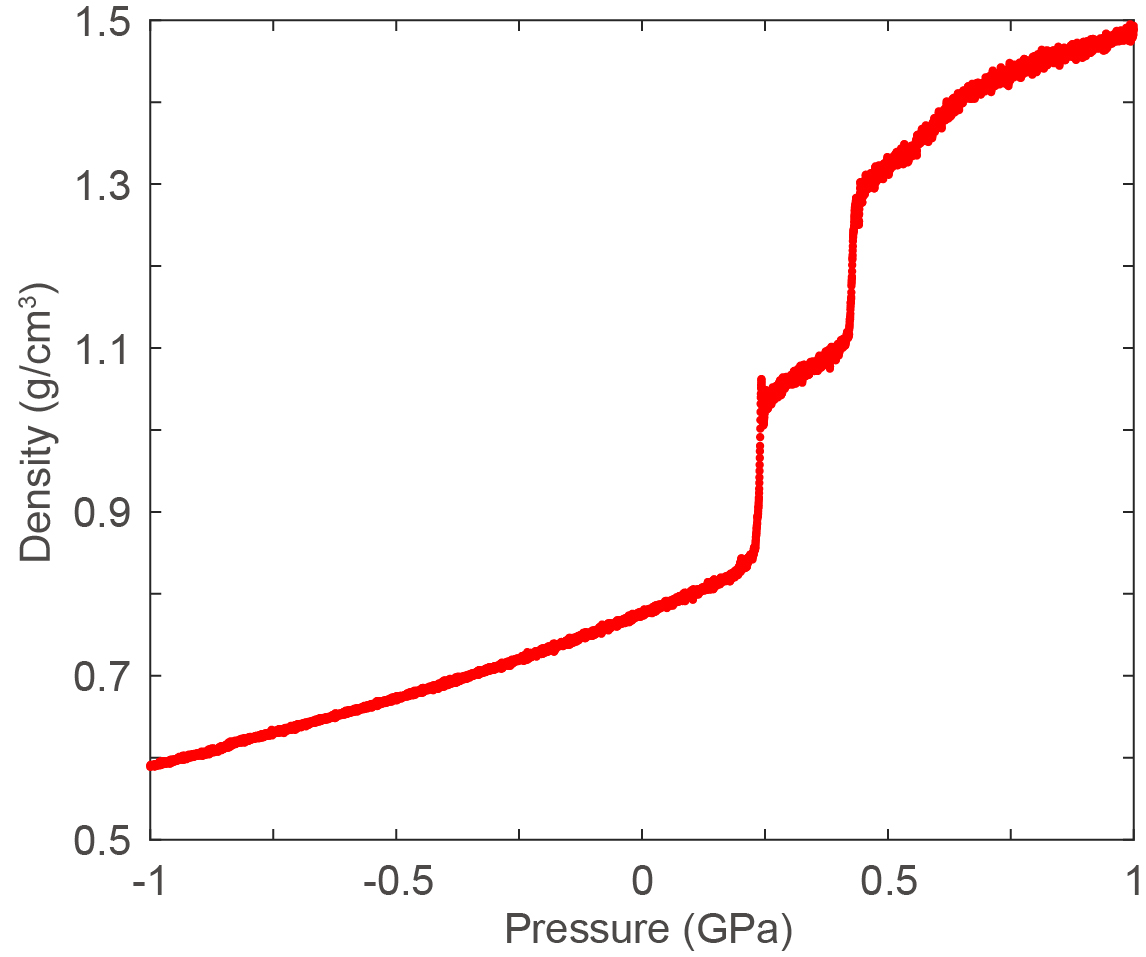}
\caption{Calculated density versus pressure ZIF-lta. Rapid increase in the density is observed at two pressures, implying pressure-induced amorphization initiated at $\sim0.25$ GPa.}
\label{SIfig:pressuredensity}
\end{figure}

In this section, we show our calculation of density versus the pressure of ZIF-lta, given in Fig.~\ref{SIfig:pressuredensity}. Following the procedure in Ref.~\cite{weng2019zifff}, we observed density drastically increases at pressures of $\sim0.25$ GPa and $\sim0.5$ GPa. According to Ref.~\cite{weng2019zifff}, initial density spike corresponds to the onset of transition from crystalline to amorphous structure.

\clearpage

\section{Acoustic phonon properties in ZIF-lta with applied strain}
\begin{figure}[hbt!] 
\includegraphics[width=\textwidth,height=.8\textheight,keepaspectratio]{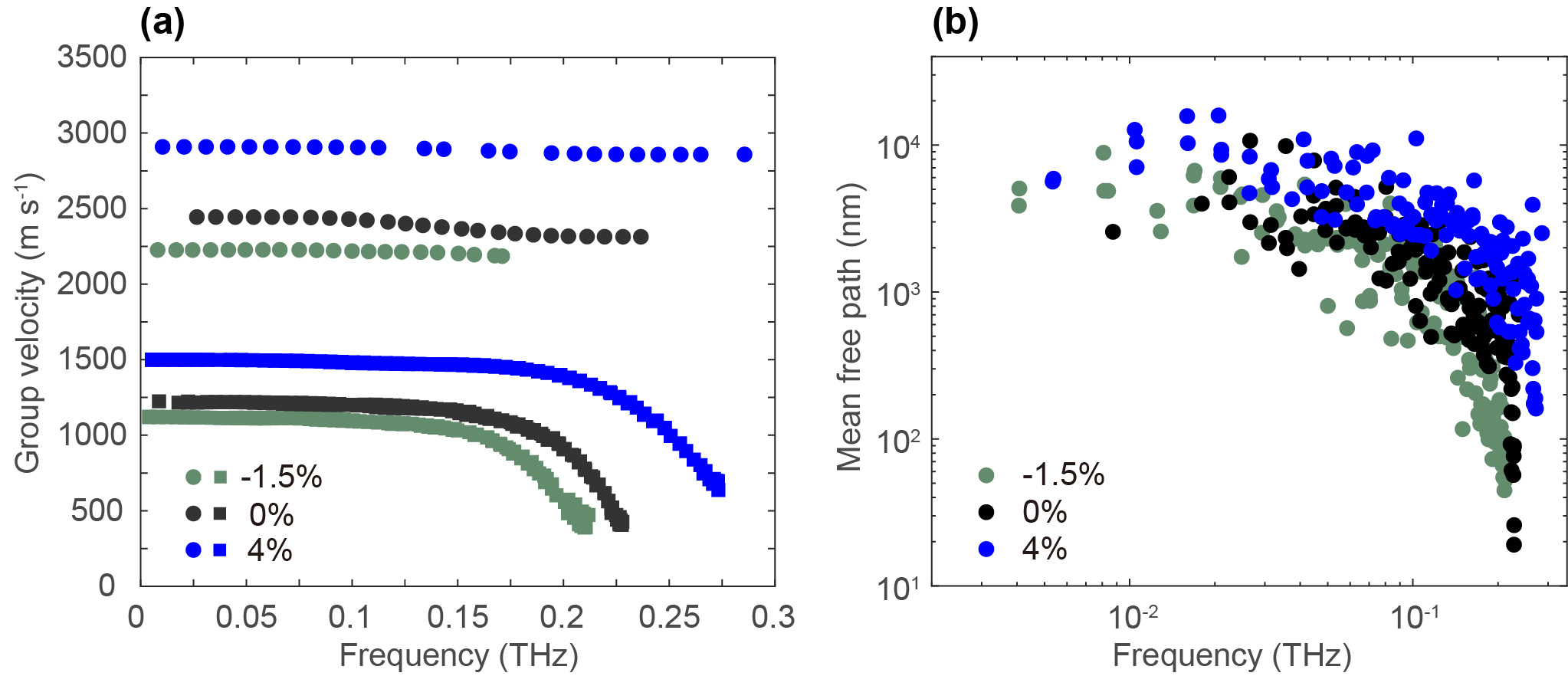}
\caption{(a) Spectral group velocities for acoustic polarizations under strains (LA modes: circles; TA modes: squares). It is observed that tensile strain increases the group velocities, while the compressive strain induces opposite effect on the group velocities. (b) Spectral MFPs for acoustic polarizations under strains. It is observed that MFPs increase from compression to expansion (-1.5\% strain: green; 0\% strain: black; 4\% strain: blue).
}
\label{SIfig:strainvg}
\end{figure}

Calculated group velocities and MFPs of acoustic modes by SED are each shown in Fig.~\ref{SIfig:strainvg}(a) and Fig.~\ref{SIfig:strainvg}(b). Specifically, the group velocities of LA modes increase from $\sim2200$ m s$^{-1}$ to $\sim2900$ m s$^{-1}$ when the applied strain changes from -1.5\% to 4\% near $\Gamma$ point. Similarly, the group velocities of TA modes increases from $\sim1100$ m s$^{-1}$ to $\sim1500$ m s$^{-1}$ when the applied strain changes from -1.5\% to 4\%. MFPs of acoustic modes generally show an enhancement under the tensile strain at a given frequency.

\clearpage

\section{VDOS comparison to bulk hydrogen gas}
\begin{figure}[hbt!] 
\includegraphics[width=\textwidth,height=.35\textheight,keepaspectratio]{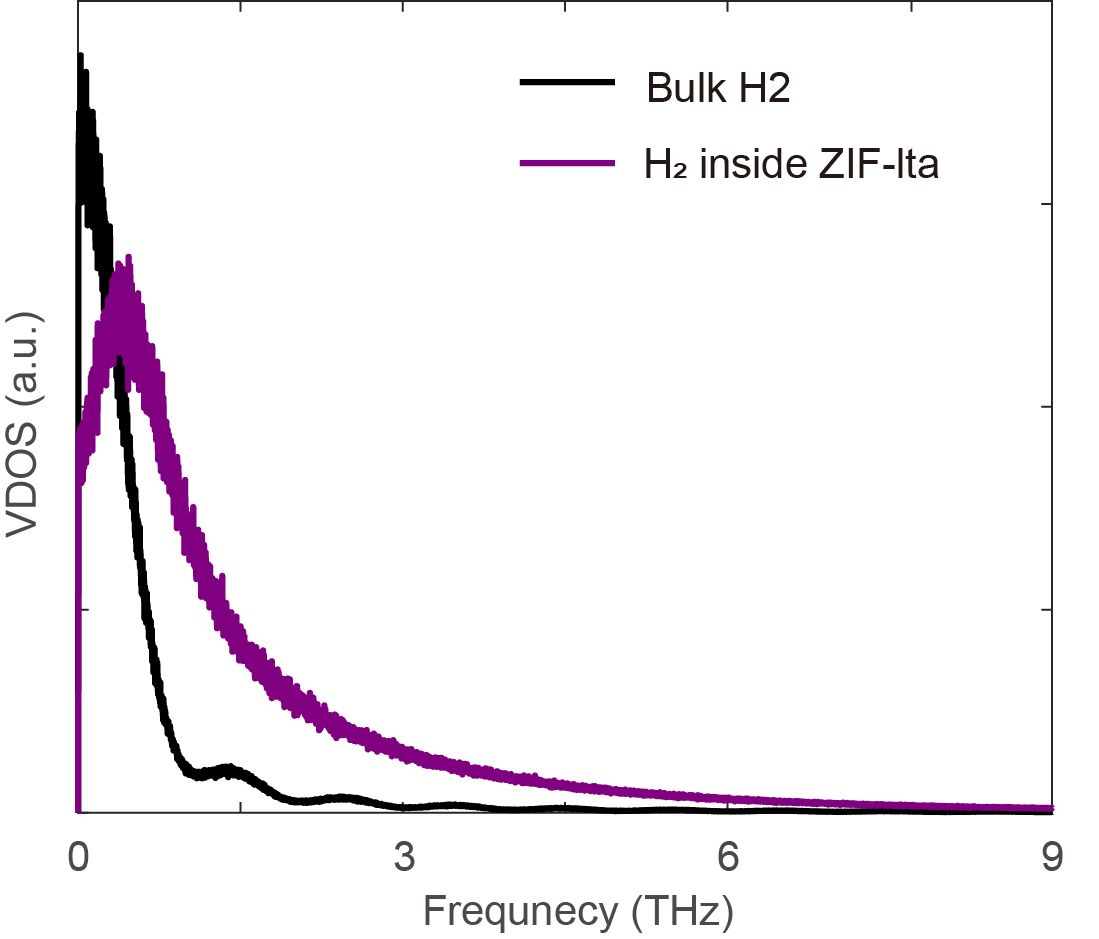}
\caption{Calculated VDOS spectra for bulk hydrogen gas molecules (black line) along with that for hydrogen infiltrated inside the structure (purple line). For both cases, density is 1.75 molecules nm$^{-3}$. VDOS of hydrogen molecules inside ZIF-lta shifts toward higher frequencies compared to bulk hydrogen molecules.} 
\label{SIfig:hydrogen_VDOS}
\end{figure}

This section provides VDOS computed for bulk hydrogen gas molecules and those for infiltrated inside ZIF-lta (density: 1.75 molecules nm$^{-3}$ for both cases), which are given in Fig.~\ref{SIfig:hydrogen_VDOS}. The VDOS spectra for hydrogen infiltrated structure is observed to be shifted toward high frequencies, compared to those for bulk hydrogen molecules. According to Ref.~\cite{thakur2025co2}, the observed shift to higher frequencies in VDOS indicates spatial confinement of hydrogen molecules in the structure, which, in turn, causes the molecules collide more frequently with themselves and with frameworks.

\clearpage

\section{Calculation of hydrogen diffusivity}
\begin{figure}[hbt!] 
\includegraphics[width=\textwidth,height=.35\textheight,keepaspectratio]{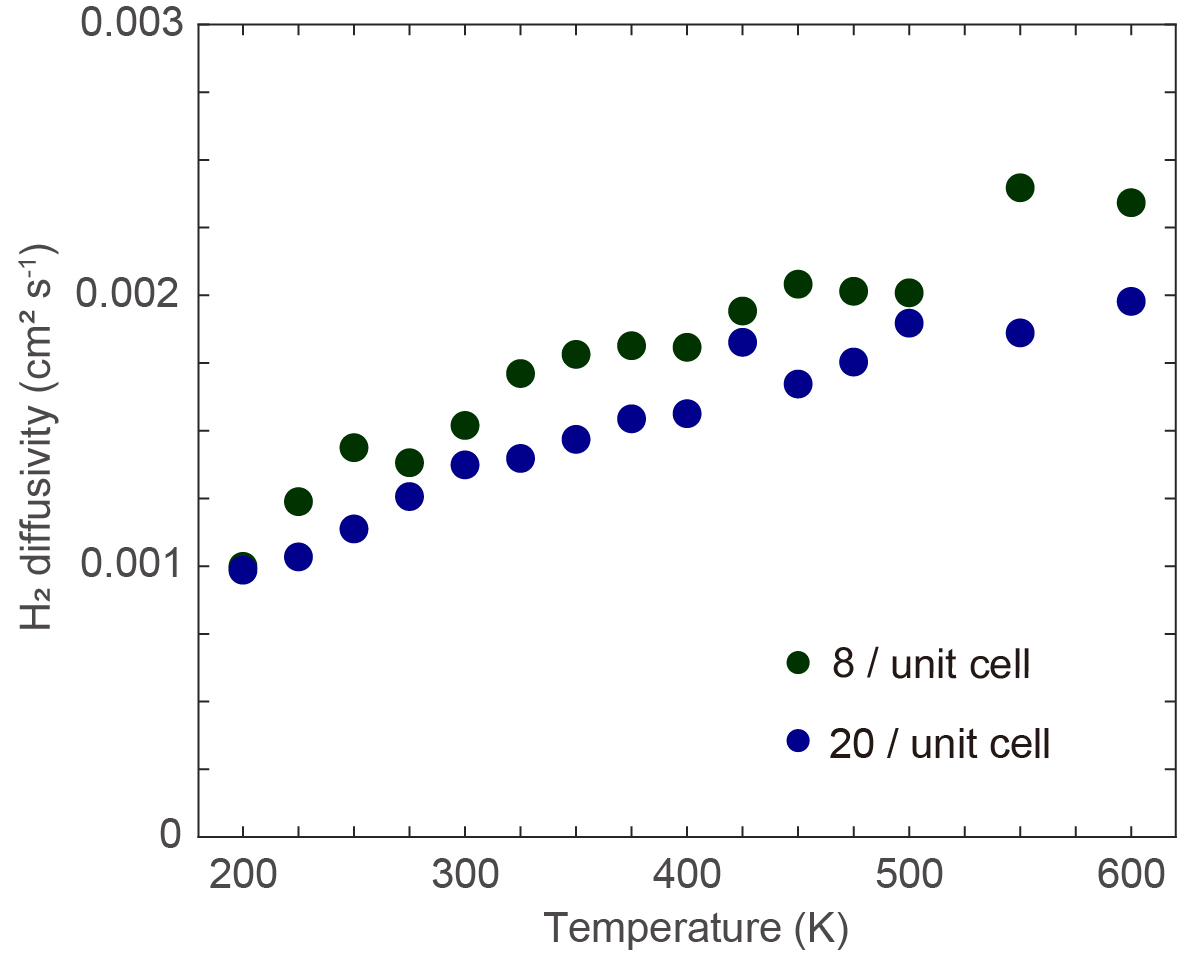}
\caption{Calculated diffusivity of hydrogen molecules infiltrated inside ZIF-lta (with 8 hydrogen molecules per unit cell: green colors; with 20 hydrogen molecules per unit cell: blue colors). Diffusivities for both cases increase as the temperature increases.}
\label{SIfig:diffusivity}
\end{figure}

In this section, we provide details of the calculation of hydrogen diffusivity. The diffusivity of the hydrogen gas was calculated using Einstein relation, expressed as
 \begin{equation}
D = \lim_{t \to \infty} \frac{1}{6N\,t} \left\langle \left|\mathbf{r_i}(t) - \mathbf{r_i}(0)\right|^2 \right\rangle,
\label{Eq:Einstein}
\end{equation}
In Eq.~\ref{Eq:Einstein}, $\left\langle \left|\mathbf{r_i}(t) - \mathbf{r_i}(0)\right|^2 \right\rangle$ is mean square displacement (MSD) of hydrogen gas molecules with total number of molecules $N$ and position $\mathbf{r_i}(t)$ at time $t$. Results are given in Fig.~\ref{SIfig:diffusivity}, showing the diffusivity that monotonically increases with temperature due to enhanced molecular mobility as the velocities of the molecules increase at higher temperatures. We note that the adsorption of 20 hydrogen molecules exhibits lower diffusivity than that of 8 hydrogen molecules, since the MSD of molecules decreases due to frequent collisions with hydrogen molecules themselves and the framework.

\clearpage

\section{Comparative analysis of the thermal conductivity of bulk hydrogen gas and that of infiltrated hydrogen in ZIF-lta}

\begin{figure}[hbt!] 
\includegraphics[width=\textwidth,height=.8\textheight,keepaspectratio]{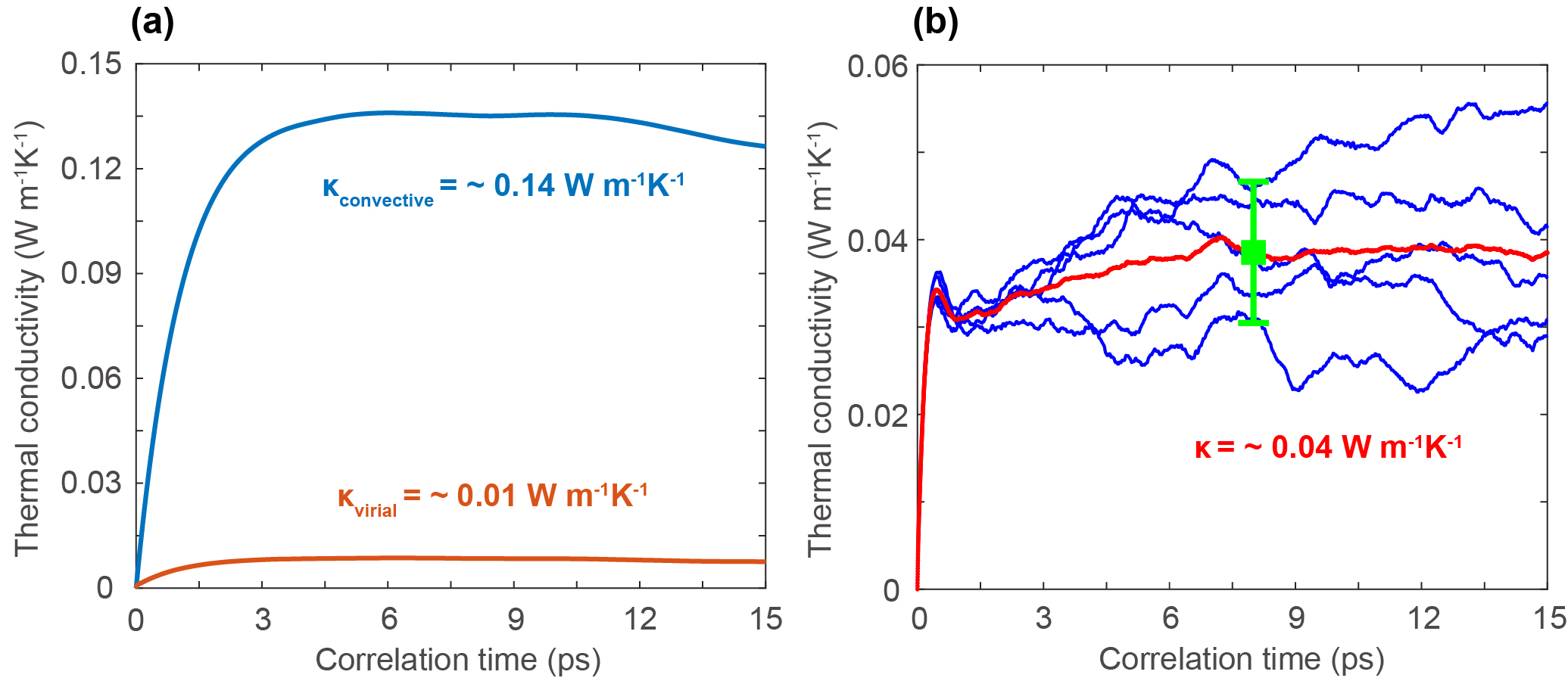}
\caption{(a) The virial and convective contributions to thermal conductivity of bulk hydrogen gas with 32 molecules per unit cell. For bulk hydrogen, it is apparent that convective contributions dominate thermal conductivity. (b) Contribution of hydrogen gas to the total thermal conductivity in infiltrated ZIF-lta with 20 molecules per unit cell. Compared to virial contribution of bulk gas, the thermal conductivity contributed by infiltrated hydrogen molecules inside ZIF-lta is clearly larger.}
\label{SIfig:k_hydrogen_combined}
\end{figure}

This section provides a comparison of thermal conductivity of bulk hydrogen molecules to that of inside ZIF-lta. For bulk case (32 molecules per unit cell) as shown in Fig.~\ref{SIfig:k_hydrogen_combined}(a), we observe that thermal conductivity is larger for convective term relative to virial term, thus the overall thermal conductivity of bulk hydrogen molecules is dominated by the convective contributions. Figure~\ref{SIfig:k_hydrogen_combined}(b) shows the hydrogen molecules contributions to the total thermal conductivity in infiltrated ZIF-lta (with 20 molecules per unit cell). Upon comparison, the virial contributions to thermal conductivity of bulk hydrogen molecules ($\sim 0.01$~\wmk) with higher density is approximately four times lower than the thermal conductivity contributed by infiltrated hydrogen molecules ($\sim 0.04$~\wmk). Noting that as discussed in the main text, the convective contribution to the overall thermal conductivity for hydrogen infiltrated ZIF-lta is negligible, the quantitative differences between the virial contribution for bulk and the thermal conductivity contribution from infiltrated hydrogen molecules indicates that gas-framework interactions predominantly contribute to enhance the apparent thermal conductivity.

\clearpage

\bibliographystyle{unsrtnat}

\bibliography{MOFRef}